# Agentic AI in Product Management: A Co-Evolutionary Model

Nishant A. Parikh

nishy.parikh@gmail.com

(Capitol Technology University, Laurel, MD 20708, USA)

## Abstract

This study explores agentic AI's transformative role in product management, proposing a conceptual co-evolutionary framework to guide its integration across the product lifecycle. Agentic AI, characterized by autonomy, goal-driven behavior, and multi-agent collaboration, redefines product managers (PMs) as orchestrators of socio-technical ecosystems. Using systems theory, co-evolutionary theory, and human-AI interaction theory, the framework maps agentic AI capabilities in discovery, scoping, business case development, development, testing, and launch. An integrative review of 70+ sources, including case studies from leading tech firms, highlights PMs' evolving roles in AI orchestration, supervision, and strategic alignment. Findings emphasize mutual adaptation between PMs and AI, requiring skills in AI literacy, governance, and systems thinking. Addressing gaps in traditional frameworks, this study provides a foundation for future research and practical implementation to ensure responsible, effective agentic AI integration in software organizations.

**Keywords**: Agentic AI, Product Management, AI-Human Collaboration, Product Lifecycle, Co-Evolutionary Model, Organizational Transformation, Product Manager Role, AI Product Manager, AI Product Development

Agentic AI in Product Management: A Co-Evolutionary Model

**Introduction**

The practice of product management has always been characterized by evolution, shifting in response to technological progress, market complexity, and customer expectations (Parikh, 2025a). From its origins in brand management at Procter & Gamble in the 1930s to its formalization in the software industry through agile and lean methodologies, the role of the product manager (PM) has expanded to encompass strategic alignment, user experience stewardship, cross-functional leadership, and finally technical and data skills required to manage AI-first products (Ebert, 2014; Parikh, 2025b).

In recent years, the integration of artificial intelligence (AI) into digital products has introduced a new layer of complexity and opportunity. Early AI applications, such as predictive analytics, recommendation engines, and natural language processing, were largely augmentative, enabling PMs to extract insights and automate repetitive tasks (Parikh, 2025b). However, the latest wave of AI technologies marks a qualitative leap: *the rise of agentic AI systems*. Figure 1 shows the exponential trends of Agentic AI in recent years (2022 to 2025). This shift reflects a broader industry transformation. According to McKinsey (2023), generative AI has the potential to add up to $4.4 trillion to the global economy by improving productivity across nearly all sectors, from customer operations and marketing to software development and research and innovation. To realize this potential, companies must learn to leverage agentic AI technologies strategically, designing products and services that align agentic AI capabilities with human goals and enterprise priorities (PwC Middle East, 2024). Product Managers, as key innovators and decision-makers, will be at the center of this transformation, responsible for guiding, integrating, and scaling these technologies in ways that maximize business value while mitigating risks (Illuri, 2025; Mariani & Dwivedi, 2024; Witkowski & Wodecki, 2025).

Agentic AI in Product Management: A Co-Evolutionary Model

**Figure 1.** Agentic AI versus Predictive AI versus AI Product Trends

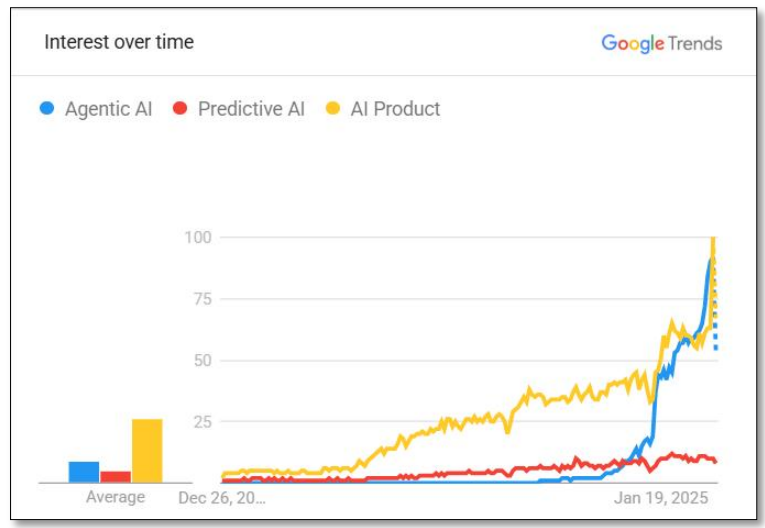

For product managers, this shift is both disruptive and transformative. Traditional product management assumes a linear relationship between customer insights, product strategy, development sprints, and releases (Cooper, 2022). In an agentic AI context, this model is challenged by AI systems that can generate product concepts, experiment autonomously, personalize features at scale, and adapt functionality in near real time (Edwards et al., 2024; Parikh, 2023). As a result, the PM's role must be reconceptualized, not as a gatekeeper of processes but as an orchestrator of intelligent ecosystems.

**Problem Statement**

The rapid advancement of artificial intelligence, particularly in its agentic AI form, is reshaping how organizations approach innovation, decision-making, and value creation. While 92% of companies plan to increase their AI investments over the next three years (McKinsey, 2025), only 1% consider themselves "mature" in AI deployment, with most organizations struggling to integrate AI meaningfully into core workflows. This gap between ambition and execution is especially pronounced in the domain of product management, where AI adoption intersects directly with strategic planning, user experience, and cross-functional coordination.

Agentic AI in Product Management: A Co-Evolutionary Model

Agentic AI represents a paradigm shift from content generation to autonomous decision-making and task execution (Acharya et al., 2025), with Deloitte forecasting that half of all enterprises will adopt such systems by 2027 (Deloitte Insights, 2024). The accelerating integration of artificial intelligence (AI) into business functions has profoundly impacted the domain of product management (Parikh, 2025b).

At the same time, failure rates for AI initiatives remain alarmingly high. Over 80% of AI projects fail to deliver expected outcomes, with primary causes including poor data quality, weak integration, and misalignment between AI outputs and business goals (Francis, 2024; Gartner, 2025; NTT Data, 2024; RAND Corporation, 2024). The specific problem is that the traditional product management frameworks are ill-equipped to accommodate systems that not only execute predefined instructions but also act with a degree of independence, adapt strategies, and evolve based on environmental stimuli (Kellogg et al., 2020; Raisch & Krakowski, 2021; Romanova, 2024). This study responds to that critical gap by addressing the absence of a conceptual model that can help product managers and software organizations understand, integrate, and govern agentic AI across the end-to-end product lifecycle.

**Purpose of the Study**

The purpose of this study is to explore, conceptualize, and articulate a framework for product management in the age of agentic AI. This research aims to (1) examine how agentic AI shifts the roles, responsibilities, and decision-making processes of product managers and (2) propose a conceptual framework that sets a foundation for researchers and practitioners to explore further and validate. Through an integrative review, the study provides a foundation for both scholars and practitioners to engage with this frontier in agentic AI product management.

## Significance of the Study

This study is significant for multiple reasons. First, it addresses a timely and underexplored dimension of agentic AI integration in organizational contexts beyond automation. Existing frameworks tend to treat AI as a tool or assistant rather than a semi-autonomous actor capable of influencing strategic direction, innovation, and execution. The study introduces a nuanced perspective on how autonomous, goal-oriented AI agents reshape the fabric of product management. Second, this research contributes significant value by offering a co-evolutionary model grounded in interdisciplinary theory that guides product managers and organizations on the future of product management and how to get ready for the significant shift. As organizations increasingly adopt generative and autonomous technologies, this study provides a critical knowledge base that informs future research and professional standards in agentic AI product management.

## Literature Review

The literature review synthesizes insights from management science, computer science, and organizational theory to provide a foundation for understanding how agentic AI is influencing the theory and practice of product management. It is structured into three subsections: (1) the evolution of product management as a discipline, (2) the integration of AI into product development processes, and (3) the emerging concept of agentic AI.

**The Evolution of Product Management**

Product management has traditionally operated at the intersection of business, technology, and user experience (Ebert, 2014). Rooted in early 20th-century brand management, the modern role evolved alongside the emergence of software products in the 1990s and the widespread adoption of agile development in the 2000s (Parikh, 2025a). The core responsibilities of product managers include identifying market opportunities, defining product strategy, roadmapping, requirements engineering, product vision creation, aligning cross-functional teams, and ensuring product-market fit (Maglyas et al., 2017).

Over the past two decades, several frameworks have shaped product management thought. The Lean Startup methodology emphasizes hypothesis-driven experimentation, minimal viable products (MVPs), and iterative learning (Ries, 2011). Agile and Scrum frameworks have formalized cross-functional team collaboration through ceremonies such as sprint planning, retrospectives, and backlog grooming (Knapp et al., 2016). Several other frameworks and bodies of knowledge provide detailed guidance on managing the entire product life cycle, from ideation to market introduction and eventual retirement (Cooper, 2022; Geracie & Eppinger, 2013; Kittlaus & Fricker, 2017).

However, these approaches assume a predominantly human-centered workflow where product managers act as facilitators and orchestrators of human cross-functional teams and decision-makers. PMs increasingly act as boundary spanners between strategic, operational, and technical domains.

**Artificial Intelligence in Product Management**

AI's entry into product management has followed a phased trajectory. Early implementations focused on predictive analytics and data mining to support decision-making. For example, AI models were used to identify churn risks, forecast demand, or optimize marketing campaigns (Agrawal et al., 2016).

The next phase involved the integration of machine learning algorithms directly into products, such as recommendation systems, dynamic pricing engines, and fraud detection mechanisms (Wang et al., 2021). These developments began to blur the boundaries between the product and the analytical system supporting it. As noted by Khare and Srivastava (2022), product managers were required to understand data pipelines, model accuracy, and algorithmic bias to make effective design and delivery decisions.

The emergence of the generative AI system marks the most recent phase in artificial intelligence development. This phase encompasses technologies such as large language models (LLMs), including GPT-4 and Claude, which can generate natural language content, write code, and synthesize documents with a high degree of fluency and contextual awareness (Achiam et al., 2023). Generative AI is reshaping the landscape of product management by enhancing every stage of the product lifecycle. Its advanced capabilities allow product managers to accelerate idea generation, conduct deeper market research, and extract actionable insights from vast quantities of customer data (Parikh, 2023).

Furthermore, agentic systems such as AutoGPT, BabyAGI, and custom LLM-based agents have been developed to autonomously plan and execute sequences of tasks with minimal human supervision. These systems are distinguished by their ability to break down complex objectives into manageable subtasks, maintain context, and adapt their actions based on new information, thereby pushing the boundaries of what autonomous AI can achieve in domains ranging from software development to market research (Acharya et al., 2025).

These tools enable new product workflows. For instance, LLMs can assist with user research by summarizing thousands of survey responses (Brand et al., 2023; Jansen et al., 2023). They can generate UX copy, wireframes, or even prototype interfaces using natural language prompts (Edwards et al., 2024). Agent-based systems can autonomously run A/B tests, iterate on features, or modify workflows based on real-time telemetry (Wang et al., 2025).

As a result, product management is becoming less about managing feature lists and more about designing, curating, and aligning agentic AI systems.

**Generative AI versus AI Agent versus Agentic AI**

*Generative AI*

Generative AI refers to a class of artificial intelligence systems, most notably large language models (LLMs) and large image models (LIMs), that are designed to generate novel outputs such as text, images, or code in response to user prompts. These models function primarily as reactive systems, producing content when prompted by a user, but they do not independently pursue objectives or interact with external environments unless explicitly directed to do so (Sapkota et al., 2025).

*AI agent*

An AI agent builds upon the capabilities of generative AI by integrating additional functionalities such as tool use, function calling, sequential reasoning, and a limited degree of autonomy. AI agents are able to perform multi-step, goal-directed tasks, retrieve real-time information, and interact with external application programming interfaces (APIs) or software systems. Despite these enhancements, AI agents typically operate as single-entity systems that execute well-defined functions within a bounded scope. They exhibit a degree of autonomy in that they can respond to user-defined goals and carry out workflows that extend beyond simple prompt-response interactions. However, their autonomy is limited to the context of a single workflow or session. AI agents thus represent an intermediate step between purely generative models and fully agentic systems, enabling automation, real-time data retrieval, and workflow execution (Sapkota et al., 2025).

*Agentic AI*

Agentic AI, in contrast, constitutes a paradigm shift toward systems composed of multiple, specialized AI agents that collaborate, communicate, dynamically perform planning, and allocate sub-tasks to achieve complex, shared objectives (Acharya et al., 2025; Hosseini & Seilani, 2025). These systems are characterized by persistent memory, dynamic task decomposition, multi-agent orchestration, and coordinated autonomy. Agentic AI systems are capable of emergent behaviors and adaptability in unstructured or dynamic environments, as they feature agents with specialized roles that can communicate and coordinate to achieve collective goals. This architecture allows for a high degree of autonomy and adaptability, enabling robust decision-making and problem-solving in complex domains. Agentic AI systems are designed to support orchestrated, real-time reasoning and adaptive control in mission-critical applications, such as research automation, robotic coordination, and advanced decision support, thereby surpassing the capabilities of single AI agents or generative models (Sapkota et al., 2025).

Please note that Generative AI, AI agents, and agentic AI should not be viewed as entirely separate concepts with rigid boundaries, and their impact on product management; instead, they represent successive stages in the evolution of AI capabilities. The focus of this study is to understand how this field of AI advancement, culminating in agentic AI, reshapes the practice and theory of product management.

## Research Methodology

This study employs Torraco's (2005) integrative review to develop a conceptual framework for understanding the co-evolutionary relationship between agentic AI and product management practices. The selection of an integrative review methodology is grounded in the need to synthesize knowledge across multiple disciplines and theoretical perspectives to address a complex, emerging phenomenon that has not been extensively studied through traditional empirical methods. The research question guiding this integrative review is, "*How is agentic AI transforming the product management discipline, and what conceptual framework can support the evolving relationship between PMs and agentic AI?*"

The literature search strategy combines database searches across four major academic databases: IEEE, ProQuest, EBSCOhost, and ACM Digital Library. Additionally, Google Scholar was used to search for articles from the other relevant sources. The search strategy employed three primary concept clusters: (1) agentic AI and autonomous systems, (2) product development and management, and (3) co-evolution and human-AI interaction.

The data extraction process systematically captured relevant information from each source using a standardized form. Both peer-reviewed articles, influential white papers, industry reports, and case study reviews were considered to ensure comprehensive coverage of conceptual developments. The analysis process employed a combination of deductive and inductive coding techniques to identify patterns, themes, and relationships within and across sources.

## Theoretical Framework

The theoretical foundation for understanding agentic AI integration in product management requires a multifaceted approach that can account for the complex, dynamic relationships between agentic AI systems and human organizational practices. This study integrates three complementary theoretical perspectives: systems theory, co-evolutionary theory, and human-AI interaction theory, to provide a comprehensive conceptual foundation that transcends the limitations of any single theoretical lens.

### Systems Theory Foundation

Systems thinking offers a powerful lens through which product managers can navigate the complexity of agentic AI integration in the product lifecycle. As an approach grounded in the study of interrelationships and dynamic feedback within complex systems, systems thinking enables managers to see beyond isolated tasks and features to recognize the broader implications of their decisions (Maani and Cavana, 2007; Senge, 2006). Unlike linear planning models, systems thinking emphasizes feedback loops and emergent outcomes that arise when autonomous AI agents interact with users, data, infrastructure, and policy environments (Meadows, 2008). In the context of agentic AI, systems thinking helps product managers anticipate second-order effects, such as how AI-driven personalization may inadvertently influence user trust, regulatory risk, or market behavior over time (Auernhammer, 2020; Forrester, 1997). Applying systems thinking in product management involves designing for adaptive governance, establishing metrics that capture long-term system health, and recognizing the organization as a living system co-evolving with AI capabilities. For example, rather than measuring success solely through feature adoption, a systems-thinking PM might evaluate how human-AI collaboration influences organizational learning, ethical compliance, and cross-functional feedback quality. As agentic AI becomes increasingly embedded in product workflows, systems thinking becomes not just a skill but a strategic necessity, enabling PMs to act as orchestrators of dynamic ecosystems rather than coordinators of static roadmaps.

This framework applies key systems theory principles, including holistic thinking, interconnectedness, feedback loops, emergence, and leverage points. These principles help to understand whole systems, recognize relationships, drive behavior and adaptation, identify emergent properties, and pinpoint where small changes can have significant effects.

**Co-Evolutionary Theory**

Co-evolutionary theory provides a dynamic lens through which to understand how human and artificial agents adapt and evolve in tandem within sociotechnical systems. Initially rooted in

Agentic AI in Product Management: A Co-Evolutionary Model

evolutionary biology, co-evolution describes the reciprocal, adaptive changes that occur when two or more entities evolve in response to each other's actions over time (Lewin & Volberda, 1999). In organizational studies, this concept has been used to explain how firms adapt alongside environmental, technological, and institutional shifts (Volberda & Lewin, 2003). When applied to agentic AI in product management, co-evolutionary theory suggests that product managers and AI systems are not static collaborators but interdependent actors whose roles, behaviors, and decision logic change in response to mutual feedback. For example, as AI agents take on increasingly autonomous responsibilities, such as user segmentation, A/B testing, or feature ideation, product managers must evolve new competencies in orchestration, supervision, and ethical delegation (Shrestha et al., 2019). Simultaneously, the AI systems themselves evolve through retraining, real-time adaptation, and user-generated feedback. This dynamic creates what scholars describe as a "fitness landscape," where both humans and machines continuously reposition themselves to maintain relevance, utility, and performance (Eisenhardt & Martin, 2000). Applying a co-evolutionary mindset in product management entails building flexible governance structures, enabling bi-directional learning, and viewing role evolution not as a threat but as a strategic opportunity for innovation and resilience. As agentic AI continues to reshape digital products and organizational workflows, co-evolution becomes a foundational concept for designing adaptive, future-ready teams and technologies.

  Co-evolutionary theory provides critical insights into how agentic AI systems and human product management practices undergo mutual adaptation over time. Unlike traditional technology adoption models that assume linear implementation processes, co-evolutionary theory recognizes that both technological and human systems possess agency and the capacity to influence each other's development trajectories (Pedreschi et al., 2024).

The co-evolutionary perspective emphasizes reciprocal adaptation, where changes in AI capabilities drive adaptations in human practices, which in turn influence AI development; temporal dynamics, recognizing that co-evolution occurs across multiple time horizons; and emergent outcomes, where the results of co-evolutionary processes cannot be predicted from understanding individual components alone (Pedreschi et al., 2024).

**Human-AI Interaction Theory**

Human–AI Interaction (HAII) theory provides a foundational framework for understanding how humans engage with artificial intelligence systems across varying levels of agency, autonomy, and task complexity. As AI systems increasingly evolve from passive tools to agentic collaborators, capable of goal-setting, reasoning, and learning, HAII theory clarifies the cognitive, behavioral, and organizational dimensions of this relationship (Amershi et al., 2019). A central tenet of the theory is mutual intelligibility; humans must understand how and why AI agents make decisions, while AI systems must align with human intent, preferences, and feedback. This alignment is especially critical in product management, where product managers orchestrate both human teams and AI systems. The quality of interaction is influenced by factors such as explainability, trust calibration, shared mental models, and control boundaries (Shneiderman, 2020).

In the context of agentic AI, these interactions become increasingly dynamic and bidirectional. AI systems now offer strategic recommendations, generate content, and may even execute decisions autonomously. At the same time, product managers supervise, intervene, and learn from these AI agents. HAII theory encourages a shift beyond traditional usability metrics, urging product managers to design systems that support effective delegation, ethical oversight, and collaborative fluency with AI (Yang et al., 2020). It emphasizes the importance of role transparency, decision traceability, and adaptive interfaces that evolve alongside user learning. As AI systems continue to self-improve and operate with greater independence, HAII theory becomes essential not just for optimizing performance but for ensuring trust, accountability, and human-centered alignment in product decision-making.

Key concepts from human-AI interaction theory include appropriate reliance, where human trust in AI systems is calibrated to match AI capabilities; mental models, which are cognitive representations that humans develop to understand and predict AI behavior; and adaptive interfaces, which evolve to accommodate changing collaboration needs and emerging use cases (Amershi et al., 2019; Sowa et al., 2021).

**The Stage-Gate Process Framework**

The Stage-Gate process framework, also known as the phase-gate process, is a structured approach to new product development that divides the journey from discovery to launch into distinct stages separated by decision gates (Cooper, 2022). The typical stages include Discovery, Scoping, Business Case Development, Development and Testing, and Launch. The study uses this framework to synthesize product life cycle, agentic AI capabilities, and their impact on product management.

## Thematic Analysis

The literature surrounding agentic AI integration in product management spans multiple disciplines yet remains fragmented and theoretically underdeveloped. This review adopts an analytical approach organized around five thematic anchors that illuminate critical gaps, theoretical tensions, and emerging patterns in the current knowledge base.

**Thematic Anchor 1: Evolution from Augmentative to Autonomous AI**

The literature reveals a fundamental conceptual shift from AI as augmentative technology to AI as autonomous agent, yet most organizational research has not adequately addressed the implications of this transformation. Traditional AI implementation literature focuses primarily on AI systems that enhance human capabilities rather than operate independently (Pedreschi et al., 2024). This augmentative perspective proves inadequate for addressing the unique challenges posed by agentic AI systems that possess significant autonomy and decision-making capabilities.

**Thematic Anchor 2: Product Management Theory and Autonomous Systems**

The product management literature could not address the implications of autonomous AI integration, with most existing frameworks assuming human-centered decision-making processes that are fundamentally disrupted by agentic AI capabilities. Recent attempts to integrate AI considerations into product management frameworks reveal the inadequacy of incremental approaches that treat AI as a tool rather than as an autonomous collaborator.

**Thematic Anchor 3: Human-AI Interaction and Governance**

The human-AI interaction literature provides essential insights into collaboration between humans and AI systems, yet most existing research focuses on assistive rather than autonomous AI applications. The trust and reliance literature offers valuable frameworks for understanding human attitudes toward AI systems, but these frameworks may not adequately address the complexity and unpredictability of autonomous AI agents.

**Thematic Anchor 4: Organizational Change and Co-Evolution**

The organizational change literature offers important insights into technology adoption and transformation processes. However, most existing frameworks assume linear implementation models that cannot accommodate the co-evolutionary dynamics of agentic AI integration. The co-evolutionary perspective, while theoretically promising, has not been adequately developed for organizational technology integration contexts.

**Thematic Anchor 5: Systems Thinking and AI Integration Complexity**

The systems thinking literature provides essential conceptual tools for understanding the complex, interconnected nature of agentic AI integration. However, these insights have not been systematically applied to product management contexts. The complexity science literature offers important insights into how complex systems adapt and evolve. However, this literature has not been systematically applied to organizational technology integration contexts.

**Critical Gaps and Synthesis**

This analytical review reveals several critical gaps in the current literature that limit understanding of agentic AI integration in product management contexts. First, the literature lacks comprehensive theoretical frameworks that can account for the autonomous nature of agentic AI systems. Second, the literature has not adequately addressed the co-evolutionary nature of human-AI collaboration. Third, the literature lacks integration across disciplinary boundaries, especially in product management. These gaps collectively highlight the need for new theoretical frameworks designed explicitly for agentic AI integration in product management contexts.

## PM-AI Co-Evolutionary Conceptual Framework

This section introduces a conceptual framework that synthesizes the findings from the literature review and theoretical frameworks. The aim is to offer a model for understanding how agentic AI transforms product management processes, roles, and organizational dynamics.

*Discovery*

The Discovery stage is fundamentally about identifying unmet market needs, understanding customer pain points, and recognizing emerging trends that can inform new product development. Traditionally, this involves extensive market research, competitive analysis, and direct customer engagement. Agentic AI can significantly enhance this stage by autonomously sifting through vast and complex datasets, identifying subtle patterns, and generating actionable insights that might otherwise be overlooked.

**Agentic AI Capabilities in Discovery.**

*Automated Market Sensing.* Agentic AI systems can continuously monitor diverse data sources, including social media, news feeds, academic publications, patent databases, and competitor announcements. They can autonomously identify nascent trends, shifts in consumer sentiment, and competitive moves, providing real-time market intelligence. This goes beyond simple data aggregation; the agents can interpret context and infer implications.

*Proactive Customer Insights*. By analyzing customer support interactions, product usage data, and public forums, agentic AI can proactively identify recurring issues, feature requests, and areas of dissatisfaction. These agents can then synthesize this information into prioritized lists of pain points or opportunities, even suggesting potential solutions based on learned patterns from successful products.

*Opportunity Identification.* Agentic AI can cross-reference market data with internal capabilities and technological advancements to identify novel product opportunities. For example, an agent might identify a gap in the market for a specific type of personalized learning tool by analyzing educational trends, available AI models, and user feedback on existing platforms.

**Impact on Product Management.**

Agentic AI in Product Management: A Co-Evolutionary Model

Product managers in the Discovery stage can transition from manual data collection and initial analysis to a more strategic role of interpreting AI-generated insights, validating hypotheses with targeted human research, and focusing on high-level strategic direction. The speed and depth of insight provided by agentic AI can significantly accelerate the discovery process and reduce the risk of pursuing non-viable ideas. Now, PMs with a systems thinking mindset create continuous feedback loops where AI agents monitor market signals, customer behavior, and competitive moves simultaneously, feeding insights back to product managers who adjust concepts in real-time.

*Illustrative Case Study 1: Airbnb*

Airbnb leverages AI agents to analyze guest feedback and identify new feature opportunities. By automating the discovery process, Airbnb's product teams can focus on high-impact initiatives and reduce time-to-market (AIM Research, 2024).

*Illustrative Case Study 2: Unilever's AI-Powered Trend Analysis*

Unilever used AI agents to scan social media, blogs, and e-commerce platforms for emerging beauty trends. Their AI system identified a surge in demand for vegan and cruelty-free skincare, leading to the successful launch of new product lines. The AI aggregated millions of unstructured data points in real time and detected signals before they became mainstream trends (Unilever, 2024). The PMs set objectives and refined what constituted "emergent demand" for the sensing agents, aligning AI insights with brand values.

*Illustrative Case Study 3: Intuit's Use of AI for Need Discovery*

Intuit leveraged LLMs to analyze customer support transcripts and automatically cluster problem areas users faced in their QuickBooks product (Intuit, 2024; AI Expert Network, 2024). This process revealed latent pain points related to tax calculation. The product team used these AI-generated clusters to simulate personas and map refined customer journeys. The PM's role was pivotal in validating these clusters and steering product planning in the right direction.

*Scoping (Solution Creation)*

Once opportunities are identified, the Scoping stage focuses on translating these into concrete product ideas, defining their core functionalities, and sketching out the user experience. This phase involves brainstorming, conceptualization, and initial prototyping. Agentic AI can act as a powerful co-creator, accelerating ideation and refining concepts with data-driven precision.

**Agentic AI Capabilities in Solutioning.**

*Generative Ideation.* Agentic AI can generate a wide array of product concepts, features, and user stories based on identified market needs and strategic objectives. These agents can draw upon vast knowledge bases of successful product patterns, design principles, and technological capabilities to propose innovative solutions. They can even generate multiple variations of a concept to explore diverse approaches.

*Automated Prototyping and UI/UX Design.* Given a set of requirements or user flows, agentic AI can autonomously generate low-fidelity prototypes, wireframes, or even functional mock-ups. These agents can adhere to design systems, ensure consistency, and iterate on designs based on simulated user interactions or predefined usability metrics. This significantly reduces the time and effort required for initial design exploration.

**Impact on Product Management.**

In the Scoping stage, product managers can shift their focus from manual ideation and design to curating AI-generated concepts, providing strategic guidance, and conducting high-level user validation. This allows for a much faster iteration cycle, enabling product teams to explore more ideas and refine promising concepts with greater efficiency and confidence. With the application of co-evolutionary theory, PMs and organizations continuously learn emergent outcomes, and simultaneously, they develop new capabilities to serve the emerging markets. The product concept evolves as both market understanding and organizational capabilities advance together. PMs now combine human creativity and strategic thinking with AI data insights and pattern recognition.

*Illustrative Case Study 4: Duolingo*

Duolingo uses generative AI to personalize lesson content and adapt difficulty in real time. The system autonomously generates exercises and feedback, reducing the manual workload for product and content teams (Marr, 2023).

*Illustrative Case Study 5: Airbnb's Sketch-to-Code Prototyping Tool*

Airbnb developed an internal tool using AI that could convert wireframe sketches into functional frontend code prototypes, dramatically accelerating design exploration (Niemeyer, 2017). This generative design tool allowed designers and PMs to experiment with multiple UX layouts in hours rather than weeks. PMs reviewed the outputs for brand alignment and user flow clarity, collaborating with AI to guide boundaries.

**Business Case Development**

The Business Case Development stage is critical for assessing the commercial viability and strategic alignment of a product concept. This involves detailed market sizing, financial modeling, risk assessment, and resource planning. Agentic AI can provide dynamic, real-time analysis and predictive modeling, offering a more robust and responsive business case.

**Agentic AI Capabilities in Business Case Development.**

***Dynamic Market Sizing and Forecasting.*** Agentic AI can continuously pull and analyze real-time market data, economic indicators, and competitor performance metrics to provide dynamic market size estimations and revenue forecasts. These agents can adjust their models based on changing conditions, offering more accurate and adaptive financial projections than static reports.

***Automated Financial Modeling and Scenario Planning.*** Given various assumptions, agentic AI can autonomously build complex financial models, conduct sensitivity analyses, and simulate multiple business scenarios (e.g., best-case, worst-case, most likely). This allows product managers to quickly understand the financial implications of different strategic choices and identify key drivers of success or failure.

**Impact on Product Management.**

Product managers in the Business Case stage can leverage agentic AI to gain more profound, more dynamic insights into product viability. This enables them to make more informed investment decisions, optimize resource allocation, and proactively address potential challenges, leading to stronger business cases and a higher likelihood of product success.

*Illustrative Case Study 6: JP Morgan COiN: A case study of AI in finance*

JP Morgan's COiN (Contract Intelligence) is a machine-learning platform designed to streamline and automate the review of legal documents, particularly commercial loan agreements, in financial transactions (Superior Data Science, 2017). Traditionally, this process was manual and prone to human error, consuming extensive resources and time. Using COiN's natural language processing and pattern recognition, JP Morgan reduced about 360,000 hours of legal review to mere seconds while improving accuracy and consistency. This transformation highlights agentic AI's ability to execute complex, specialized tasks with a high degree of autonomy, freeing Product Managers and financial teams to focus on more strategic responsibilities, developing and validating new financial products, pricing strategies, and risk models instead of getting bogged down in routine paperwork. Ultimately, COiN underscores how agentic technology can drive operational efficiency and help financial institutions accelerate their product development lifecycle.

### *Developing and Testing*

The Development and Testing stage encompasses the comprehensive process of building, testing, and refining the product through iterative cycles of development, quality assurance, and validation. This stage involves software development, system integration, comprehensive testing protocols, user validation, and continuous refinement based on technical and user feedback. Agentic AI can fundamentally transform this stage by enabling autonomous development assistance, intelligent testing orchestration, and continuous optimization throughout the development lifecycle.

**Agentic AI Capabilities in Development and Testing:**

*Autonomous Code Generation and Development Assistance:* Agentic AI can actively participate in the development process by generating code components, suggesting architectural improvements, and identifying potential technical debt. These agents can analyze requirements, automatically generate unit tests, implement feature specifications, and even refactor existing code for better performance and maintainability. They can conduct real-time code reviews to verify compliance with coding standards and best practices, thus improving development speed.

*Intelligent Test Orchestration and Automated Quality Assurance:* Agentic AI can design, execute, and manage comprehensive testing strategies across multiple environments and platforms. These agents can generate test cases, perform regression, load, and stress testing, and assess security vulnerabilities. They also manage complex testing workflows, handle test data, and prioritize tests based on risk and code changes.

***Continuous Integration and Deployment Optimization:*** Agentic AI can manage and optimize the entire CI/CD pipeline, automatically detecting build failures, resolving dependency conflicts, and ensuring seamless integration across development teams. These agents can monitor code quality metrics, automatically trigger appropriate testing protocols, and manage deployment strategies while maintaining system stability and minimizing deployment risks.

**Impact on Product Management:** Product managers in the Development and Testing stage can leverage agentic AI to accelerate development cycles, ensure higher quality standards, and maintain continuous alignment between technical implementation and user needs. This enables them to focus on strategic product decisions, feature prioritization, and stakeholder alignment while ensuring that development and testing processes are optimized for efficiency, quality, and user satisfaction. The integration of agentic AI changes development management into a data-driven process that can adjust to evolving requirements and market conditions in real time. Now, PMs with systems thinking treat development as a system where engineering, design, marketing, and business teams are interconnected components; understand that product features emerge from team interactions, and focus on optimizing the whole development system rather than individual components.

*Illustrative Case Study 7: Microsoft GitHub Co-pilot in Agile Workflows*

GitHub Co-pilot, developed by OpenAI and Microsoft, assists developers by auto-suggesting code and boilerplate during development (Microsoft, 2025). Teams at Microsoft integrated Co-pilot into their Agile sprint planning, using it to accelerate iteration cycles. AI agents were even used to flag tickets that could be automated, freeing developer time. PMs ensured that auto-sprint agents aligned with team goals and maintained adherence to coding standards and ethical usage policies.

Agentic AI in Product Management: A Co-Evolutionary Model

*Launch*

The Launch stage involves bringing the product to market, monitoring its performance, and managing post-launch activities. This includes deployment, performance monitoring, and customer support. Agentic AI can ensure a smoother launch and continuous post-launch optimization.

**Agentic AI Capabilities in Launch.**

***Automated Deployment and Release Management.*** Agentic AI can orchestrate complex deployment pipelines, ensuring seamless and error-free releases. These agents can manage infrastructure provisioning, code deployment, and configuration, reducing manual errors and accelerating time-to-market. They can also manage phased rollouts to specific user groups.

***Proactive Performance Monitoring and Incident Response.*** Agentic AI systems can continuously monitor product performance, system health, and user experience metrics in real time. Upon detecting anomalies or potential issues, these agents can autonomously diagnose root causes, initiate automated remediation actions (e.g., scaling resources, restarting services), and escalate to human teams only when necessary. This minimizes downtime and ensures high availability.

***Continuous Post-Launch Optimization.*** Agentic AI can analyze post-launch data, including user engagement, conversion rates, and monetization metrics. These agents can then autonomously identify areas for improvement, suggest optimizations (e.g., pricing adjustments, content recommendations), and even implement small-scale changes to maximize product success and user retention. This transforms product management into a continuous optimization loop.

**Impact on Product Management.** In the Launch stage, product managers can rely on agentic AI for robust and proactive post-launch management. This frees them to focus on strategic growth initiatives, market expansion, and long-term product vision, knowing that the product is being continuously monitored and optimized by intelligent autonomous systems. With a systems thinking mindset, PMs implement feedback loops at the user level, market level, and ecosystem level to continuously monitor emerging outcomes and enhance products accordingly.

*Illustrative Case Study 8: Microsoft Azure - Agentic DevOps with GitHub Co-pilot*

Microsoft is applying agentic AI to DevOps by integrating GitHub Co-pilot with Azure to streamline and accelerate software delivery. The agent assists engineers by suggesting code, automating routine tasks, and orchestrating delivery workflows, reducing manual effort and bottlenecks. It evolves alongside the codebase and team practices, allowing Product Managers to shorten delivery cycles, respond faster to changing requirements, and innovate more rapidly. By acting as a collaborative counterpart to human engineers, this approach highlights the potential for agentic AI to transform the developing stage of the Product Lifecycle, freeing Product Managers and technical teams to focus on higher-value initiatives (Microsoft, 2025).

*Illustrative Case Study 9: Bank of America's virtual assistant Erica*

Bank of America's virtual assistant Erica exemplifies the practical application of agentic AI in financial services. Since its launch, Erica has facilitated over 2 billion interactions, supporting 42 million clients with banking, budgeting, and financial decision-making (Bank of America, 2024). Its success stems from combining natural language processing with real-time data insights, allowing it to anticipate user needs and deliver personalized assistance. This case highlights how AI can scale customer support, improve financial literacy, and enhance user experience in high-volume service environments while maintaining trust and usability across diverse demographics.

**Relationships: A Co-Evolutionary Model**

The integration of agentic AI across the product lifecycle introduces a dynamic and intricate interplay between evolving agentic AI capabilities and the evolving responsibilities of product managers. This relationship is far from linear or mechanistic. Instead, it is characterized by continuous adaptation, reciprocal influence, and emergent complexity, akin to a co-evolutionary model, as shown in Figure 2.

**Figure 2. PM-AI Co-Evolutionary Model**

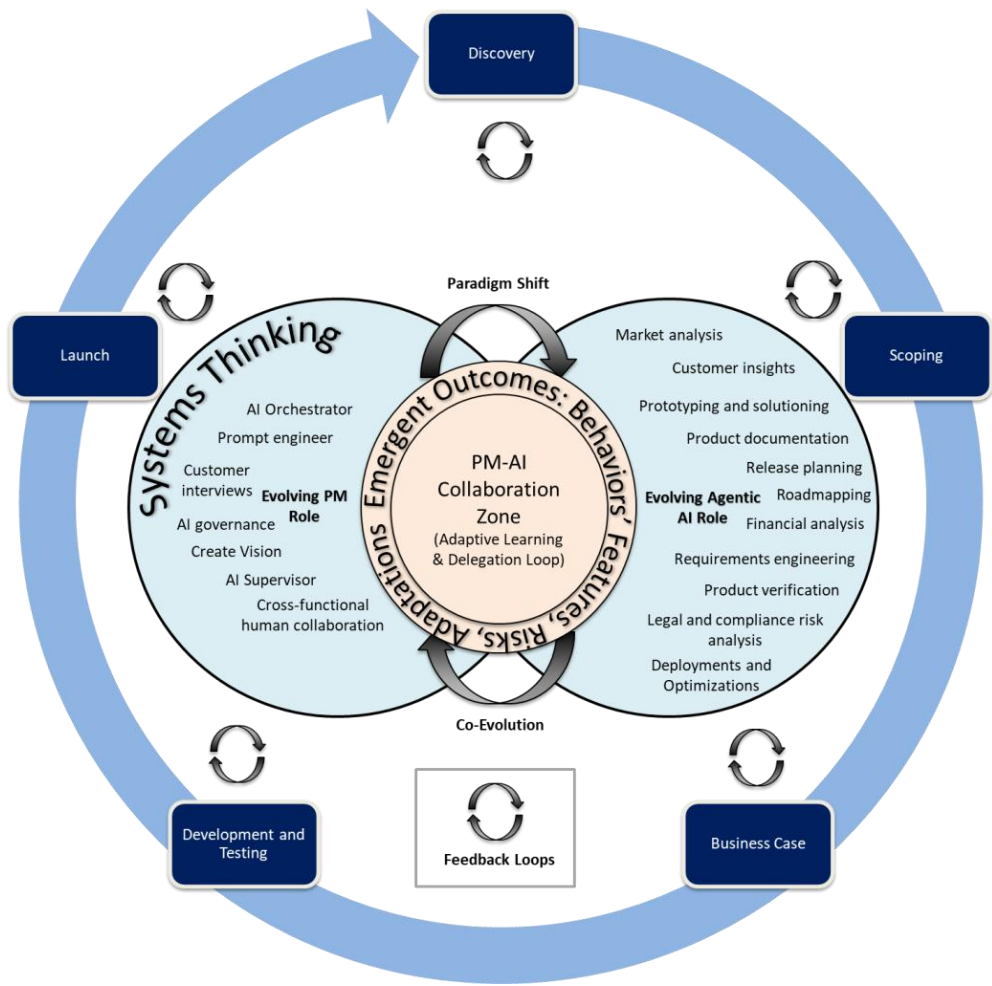

Agentic AI in Product Management: A Co-Evolutionary Model

First, a co-evolutionary relationship defines how product managers and agentic AI mutually shape each other's behavior and capabilities. As AI agents become increasingly capable of autonomous task execution, the PM's role demonstrably shifts toward higher-order functions. These include meticulous ethical supervision, sophisticated prompt engineering, the strategic alignment of multi-agent outputs with overarching organizational goals, and the critical interpretation of AI-generated insights. Simultaneously, the design, training, and refinement of AI agents are profoundly shaped by human guidance, feedback, and the constraints established by PMs, creating a continuous feedback loop of mutual learning and refinement. PMs learn how to leverage AI better, and AI systems learn to serve product goals better.

Second, the integration of agentic AI transforms traditional team structures into complex sociotechnical ecosystems in which human and machine actors collaborate, sometimes seamlessly, sometimes with friction. Product managers evolve from being primarily facilitators of cross-functional human teams into sophisticated orchestrators of hybrid workflows. In these new workflows, AI agents may act independently, cooperatively with human team members, or even, if not correctly designed or governed, in ways that appear adversarial or counterproductive to how objectives are encoded. The PM must now ensure that AI agents align rigorously with both organizational values and user-centered outcomes. This necessitates an expanded toolkit for governance, continuous monitoring, transparent communication, and timely intervention when AI actions deviate from desired paths.

Third, this evolving relationship is governed by an increasing and critical emphasis on governance, accountability, and transparency. As agentic systems take on tasks with significant strategic implications, such as feature prioritization, financial modeling, customer segmentation, or even direct user interaction, PMs become increasingly accountable for the traceability, fairness, reliability, and ethical implications of those AI-driven decisions. This necessitates the adoption and enforcement of robust practices in model validation, ongoing performance monitoring, human-in-the-loop supervision where appropriate, and precise documentation of AI decision-making processes. Defining accountability for AI errors or unintended consequences is a significant challenge.

Finally, the framework reflects a fundamental paradigm shift in decision-making logic. Whereas traditional product management often assumes a command-and-control structure where human decision-makers initiate actions, and AI systems execute them, agentic AI can reverse and significantly complicate this hierarchy. AI agents may now autonomously generate hypotheses, propose experiments, identify market opportunities, or even initiate user-facing changes. PMs must, therefore, act less as gatekeepers of decisions and more as designers of behavioral guardrails, architects of ethical frameworks, and shapers of the conditions under which AI agents operate, learn, and adapt. Their role becomes one of enabling responsible autonomy rather than direct control.

In essence, the relationship between agentic AI and the evolving PM role is recursive, deeply adaptive, and highly contextual. It demands a new conceptualization of agency itself, not as a binary trait held exclusively by either humans or machines but as a distributed function of well-aligned goals, robust ethical oversight, transparent processes, and truly collaborative intelligence. This evolving interdependence underscores the urgent need for future research to develop empirically grounded models of PM-AI agent collaboration, adaptive governance structures, and outcome-centered performance metrics specifically tailored for AI-augmented product environments.

**Discussion**

This conceptual review set out to explore how agentic AI is being applied across the product lifecycle and how product management roles are evolving in response. The proposed framework, derived from a synthesis of multidisciplinary literature and theoretical frameworks, reveals several key insights and implications for both research and practice.

First, the analysis shows that agentic AI is a fundamental shift in product ideation, conception, development, launch, and optimization, not just an incremental improvement over previous AI tools. Agentic AI systems, with their autonomous goal-setting, adaptive learning, and multi-agent collaboration, are now capable of performing tasks previously done by human teams. In the discovery phase, agentic AI can autonomously analyze market trends and generate novel product concepts, accelerating discovery and reducing bias. In design and development, AI-driven prototyping and iterative testing enable rapid experimentation and refinement, while in launch and growth, agentic systems optimize campaigns, integrate feedback, and personalize user experiences in real-time.

Second, the evolving role of the product manager emerges as a central theme. PMs are transitioning toward higher-level responsibilities such as curating and interpreting AI-generated insights, orchestrating AI agents, ensuring ethical oversight, and governing the deployment and scaling of agentic systems. This shift demands new competencies in AI literacy, prompt engineering, model governance, and ethical stewardship, as well as the ability to design and supervise complex sociotechnical ecosystems. The framework highlights that successful integration of agentic AI hinges not only on technical adoption but also on organizational readiness, cross-functional alignment, and a commitment to responsible innovation.

Agentic AI in Product Management: A Co-Evolutionary Model

Expanding upon this, Raisch and Krakowski's (2021) paradox perspective on automation and augmentation highlights a crucial consideration for Product Managers navigating this transformation. Raisch and Krakowski argue that automation and augmentation are not separate or opposing strategies but two intertwined aspects of a paradox that must be managed across time and context. Instead of choosing between letting machines operate independently or keeping humans "in the loop," Product Managers need to learn to embrace and balance both, designing products and organizations that maximize synergy while mitigating drawbacks.

This view underscores the necessity for new management practices and perspectives. Product Managers must develop the judgment to determine when to allow an AI to execute tasks autonomously and when to employ human expertise to augment its capabilities. Importantly, this approach highlights their role as key players in navigating paradox and shaping collaboration, making decisions about when, where, and how PM and AI components should be blended.

Together, these perspectives illuminate a future in which Product Managers become orchestrators of AI agents responsible for designing and governing a dynamic ecosystem of roles, responsibilities, and interactions. To perform this role effectively, Product Managers must move away from rigid, linear methods and embrace adaptive, paradox-aware practices that align human judgment and algorithmic power in service of a shared mission.

Additionally, the rise of agentic Artificial Intelligence underscores a growing need for Product Managers to become responsible innovators and policy advocates (Murugesan, 2025). This view highlights that agentic AI systems, defined by their ability to pursue goals, learn from their environment, and make decisions with a degree of independence, bring new opportunities alongside significant responsibilities. As Product Managers integrate these autonomous components into products and services, they must establish robust governance mechanisms, ethical safeguards, and oversight structures to align agent behavior with corporate goals, legal standards, and societal values. This means Product Managers are not only designing for functionality and market fit but also designing for responsible action, proactively addressing issues of fairness, transparency, and accountability. In this context, Product Managers become key stakeholders in shaping policy, guiding implementation, and ensuring that the rise of agentic AI delivers benefits while mitigating potential harms, reflecting a transformation in their role from innovators to responsible innovators.

## Practical Implications and Applications

This framework helps both scholars and practitioners understand and manage the integration of AI into product management. It outlines an emerging shift from traditional task-oriented roles toward collaborative and supervisory engagements between humans and autonomous systems. The framework provides a roadmap for empirical exploration into the operational, strategic, and ethical dimensions of agentic AI within product ecosystems.

For practitioners, the model highlights the growing need for upskilling, organizational restructuring, and ethical alignment. Product Managers (PMs) must now acquire new competencies in AI supervision, model governance, prompt engineering, AI agent orchestration, and most importantly, a systems thinking mindset. These skills are essential to ensure the responsible and effective deployment of AI agents throughout the product lifecycle.

For product managers, implementing agentic AI effectively requires adopting a systems perspective, viewing the product not in isolation, but as embedded within broader stakeholder, organizational, and market systems. Product managers should also embrace co-evolution, allowing their product strategies and organizational capabilities to develop in tandem with AI-driven transformations. In parallel, they must develop partnership skills that enable them to work collaboratively with AI agents, not simply as tools, but as adaptive co-creators.

For organizations, successful integration of agentic AI depends on building system-level capabilities. This includes investing in infrastructure that supports seamless human–AI collaboration, from data pipelines to orchestration platforms. In addition, organizations should create learning mechanisms that enable continuous adaptation and mutual evolution between teams and technologies. Governance frameworks must also be developed to manage risks, ensure ethical compliance, and clarify decision boundaries between humans and AI systems.

In the longer term, as agentic AI systems advance toward capabilities akin to artificial general intelligence (AGI), the current co-evolutionary relationship between human product managers and AI may shift toward an increasingly asymmetric dynamic. Rather than maintaining a balanced partnership, the trajectory could lead to agentic systems autonomously managing the entire spectrum of product management functions, including market analysis, strategic planning, prototyping, development, launch execution, and continuous optimization, without the need for sustained human oversight. In this scenario, the role of the human product manager may become marginal, confined to occasional strategic input or ethical governance, and potentially rendered obsolete altogether. Such a shift would represent a significant departure from the collaborative vision of co-evolution, signaling a future where AI capabilities continue to progress while human roles risk becoming static or diminished.

This transformation has broader implications beyond product management, raising critical questions about the future of human judgment, creativity, and leadership in increasingly autonomous, AI-driven organizational ecosystems. Notably, leading technology firms, including Google, Amazon, Microsoft, IBM, Meta, and Salesforce, have already begun restructuring their workforces, citing AI-enabled productivity gains and the consolidation of job functions as key drivers (Napolitano, 2024; Westfall, 2025; Shibu, 2025; Dellinger, 2025; Mihov, 2025; CNBC, 2025). These developments suggest a short-term trend toward leaner organizations staffed by hybrid-skilled professionals who collaborate closely with agentic systems. However, the long-term trajectory remains uncertain and demands continued exploration to understand the evolving interplay between human capabilities and increasingly autonomous AI.

## Future Research Directions

The co-evolutionary framework opens numerous avenues for future research that can advance both theoretical understanding and practical application of agentic AI integration in organizational contexts.

### Empirical Validation and Testing

The most immediate research priority involves comprehensive empirical validation of the co-evolutionary framework across diverse organizational contexts. Longitudinal studies tracking organizations through multiple phases of agentic AI integration would provide valuable insights into the framework's predictive validity and practical utility.

### Extension to Other Organizational Contexts

While the co-evolutionary framework represents a significant advancement, several limitations must be acknowledged. The framework's development is based primarily on conceptual synthesis and has no empirical validation. The framework's focus on technology-intensive software organizations limits its generalizability.

Future research should explore how the co-evolutionary framework applies to organizational functions beyond product management and software organizations. Studies examining agentic AI integration in marketing, operations, human resources, health care, education, and manufacturing would test the framework's generalizability.

**Operational Frictions and Decision Conflicts**

A key area for future investigation is the strategic and operational tension between human judgment and AI-driven recommendations. As agentic AI systems become capable of proposing product features, timelines, or prioritization strategies, researchers must examine how organizations resolve conflicts when AI output contradicts human intuition, stakeholder input, or real-time market feedback. These conflicts raise critical governance challenges: Who has ultimate authority in decision-making? How is trust established and monitored between PMs and AI agents? What oversight structures are needed when AI operates across cross-functional domains without human-in-the-loop intervention?

Empirical studies using real-world case examples, such as failed launches, ignored AI insights, or misaligned prioritization, could yield valuable insights into how friction manifests and is managed. In particular, longitudinal case studies from early adopting firms (e.g., Google, Microsoft, Intuit, Salesforce, IBM) could reveal how product teams negotiate power-sharing with AI, revise workflows, and adapt feedback loops over time.

**Ethics, Regulatory, and Governance Challenges**

Future research should prioritize the development and empirical validation of robust ethical and governance frameworks tailored for agentic AI in product management. As agentic AI systems assume greater autonomy in decision-making, critical questions arise about accountability, fairness, and transparency in their actions. Studies could explore specific governance mechanisms, such as real-time auditing tools, bias detection algorithms, and human-in-the-loop protocols, to ensure alignment with organizational values and regulatory standards.

The regulatory landscape surrounding AI-enabled product decisions remains fluid and underdeveloped. As agentic AI begins to affect market offerings, pricing, targeting, and even ethical compliance, future research should address how legal and policy frameworks will interact with autonomous systems. For instance, how can organizations maintain compliance with AI regulations like the EU AI Act or U.S. algorithmic accountability policies when AI decisions evolve continuously? Who is liable when autonomous agents make high-stakes decisions in fraud detection, healthcare, or financial services?

**Advanced Theoretical Development**

Future theoretical development should explore the intersection between co-evolutionary dynamics and emerging AI capabilities. As AI systems become more sophisticated and autonomous, new theoretical frameworks may be needed to understand advanced forms of PM-AI collaboration.

## Conclusion

This study presents a conceptual framework for understanding the evolving relationship between agentic AI and product management, grounded in systems thinking, co-evolutionary theory, and human–AI interaction. By mapping agentic AI capabilities across the product lifecycle, the framework illustrates how these systems are transforming not only product development processes but also the strategic, ethical, and supervisory responsibilities of product managers. Rather than being displaced, product managers are emerging as orchestrators of complex, adaptive ecosystems that integrate human judgment with machine autonomy.

The findings highlight that agentic AI is no longer a mere tool for efficiency but a collaborative force reshaping how products are discovered, designed, developed, launched, and optimized. Product managers must develop new competencies, particularly in AI orchestration, ethical oversight, and systems governance, to navigate this landscape responsibly and effectively. The proposed co-evolutionary model emphasizes the mutual adaptation between humans and AI, where both systems evolve in tandem to achieve strategic alignment and organizational learning.

This research contributes a foundational perspective for scholars and practitioners, offering a roadmap to guide future empirical inquiry and practical implementation. Moving forward, there is a critical need to validate the model in real-world contexts, extend it to other functional domains, and explore the long-term implications of increasingly autonomous AI on organizational structure and human roles. By fostering responsible innovation and collaborative intelligence, product management can help ensure that agentic AI systems enhance human creativity, judgment, and societal well-being.

Agentic AI in Product Management: A Co-Evolutionary Model

Agentic AI in Product Management: A Co-Evolutionary Model

Agentic AI in Product Management: A Co-Evolutionary Model

Agentic AI in Product Management: A Co-Evolutionary Model